\documentstyle[12pt,epsf,ws]{article}
\newdimen\rotdimen
\def\vspec#1{\special{ps:#1}}
\def\rotstart#1{\vspec{gsave currentpoint currentpoint translate
   #1 neg exch neg exch translate}}
\def\rotfinish{\vspec{currentpoint grestore moveto}}
\def\rotr#1{\rotdimen=\ht#1\advance\rotdimen by\dp#1%
   \hbox to\rotdimen{\hskip\ht#1\vbox to\wd#1{\rotstart{90 rotate}%
   \box#1\vss}\hss}\rotfinish}
\def\rotl#1{\rotdimen=\ht#1\advance\rotdimen by\dp#1%
   \hbox to\rotdimen{\vbox to\wd#1{\vskip\wd#1\rotstart{270 rotate}%
   \box#1\vss}\hss}\rotfinish}%
\def\rotu#1{\rotdimen=\ht#1\advance\rotdimen by\dp#1%
   \hbox to\wd#1{\hskip\wd#1\vbox to\rotdimen{\vskip\rotdimen
   \rotstart{-1 dup scale}\box#1\vss}\hss}\rotfinish}%
\def\rotf#1{\hbox to\wd#1{\hskip\wd#1\rotstart{-1 1 scale}%
   \box#1\hss}\rotfinish}%

\begin{document}
\title{ Meson and Quark Degrees of Freedom and the Radius of the
Deuteron}
\bigskip
\bigskip
\author{
A. J.  Buchmann\thanks{ Present Address: Institute of Theoretical Physics, 
University of T\"ubingen, Auf der Morgenstelle 14, D-72076 T\"ubingen, 
Germany } , H. Henning,  and P. U. Sauer  \\
Institute for Theoretical Physics \\
University of Hannover \\
Appelstra\ss e 2  \\
D-30167 Hannover \\
Germany }
\medskip
\bigskip
\maketitle
\bigskip
\begin{abstract}
\noindent
The existing experimental data for the deuteron charge radius are
discussed. The data of elastic electron scattering are inconsistent with
the value obtained in a recent atomic physics experiment. Theoretical 
predictions based on a nonrelativistic description of the deuteron with 
realistic nucleon-nucleon potentials and with a rather complete set of 
meson-exchange contributions to the charge operator are presented. 
Corrections arising from the quark-gluon substructure of the nucleon are 
explored in a nonrelativistic quark model; the quark-gluon corrections, 
not accounted for by meson exchange, are small. Our prediction for the 
deuteron charge radius favors the value of a recent atomic physics experiment.
\end{abstract}
\noindent
PACS number(s): 21.45.+v, 21.10.Ft, 25.30.Bf, 27.10+h, 12.39.Pn

\baselineskip 14pt
\noindent

\section {Introduction} 
\nobreak \noindent 
Deuteron properties are quantities of fundamental importance. The hadronic 
properties are used to calibrate parametrizations of the nucleon-nucleon (NN) 
interaction. The electromagnetic (e.m.) properties are crucial tests of the 
theory of e.m. currents, but simultaneously also of the chosen form of the NN 
interaction. Although there has been a continuous interest in all deuteron
observables, the observable root-mean-square (rms) charge radius $r_{ch}$ and
the related but unobservable rms matter radius $r_m$ have recently received
special attention \cite{Kla86,Ker91,Bha90,Spr90,Dij89,Shk94,Ker90,Won92}.
This is mainly due to unresolved discrepancies between experimental data and
theoretical predictions. These discrepancies have even led to speculations 
\cite{Spr90,Ker90,Won92} that quark-gluon effects may surface in the deuteron 
charge radius. We are sceptical that these effects can resolve the observed
discrepancies. Nevertheless, possible quark-gluon effects in the deuteron
charge radius deserve a careful investigation. This is the reason why this
paper studies general short-range corrections of hadronic and of quark-gluon
nature for the deuteron charge radius, not discussed before. With respect
to the history of the subject, Ref. \cite{Won94} gives a comprehensive review
of both experimental and theoretical aspects of the deuteron radius.
\par
Sect. 2 summarizes the existing experimental data of elastic electron 
scattering and of an atomic physics experiment relevant for the deuteron
charge radius. It does not contain novel material. Despite the review of
Ref. \cite{Won94} the section is important for this paper, since it sets the 
stage for our 
calculations; it extends our discussion in Ref. \cite{Sau94}. All calculations
of deuteron charge properties require meson-exchange (MEC) corrections for the
charge operator. Sect. 3 presents our theoretical predictions. Subsect. 3.1 
discusses them in a hadronic description of the deuteron. Subsect. 3.2 extends
this description by taking the quark-gluon substructure of nucleons into 
account. We give our conclusions on the unresolved discrepancy between 
experimental data and theoretical predictions in Sect. 4.

\section {Experimental Data and their Analyses}
\noindent
The mean square (ms) deuteron charge radius $r_{ch}^2$ is defined as the
slope of the deuteron charge monopole form factor $F_{ch}({\bf Q}^2)$ of 
elastic electron scattering at zero three-momentum transfer ${\bf Q}$, i.e.,

\begin{equation}
\label{rdf}
r_{ch}^2 = -6 {d \over {d\,{\bf Q}^2}}F_{ch}({\bf Q}^2)\vert_{{\bf Q}^2=0}.
\end{equation}

\noindent
Equivalently, it can be derived from the slope of the deuteron longitudinal
structure function $A({\bf Q}^2)$ at zero momentum transfer, corrected for a
small contribution arising from the deuteron magnetic moment $\mu_d$ measured
in units of nuclear magnetons $\mu_N=e/2M_p$, where $M_p$ is the proton mass,

\begin{equation}
\label{rda}
r_{ch}^2 = -3 {d \over {d\,{\bf Q}^2}}A({\bf Q}^2)\vert_{{\bf Q}^2=0} 
+ {\mu_d^2 \over {2 M_p^2}} .
\end{equation}

\noindent
The latter correction $\mu_d^2 / 2 M_p^2 = 0.0163$ fm$^2$ arises, since the
longitudinal structure function $A({\bf Q}^2)$ is built up from the charge 
monopole, charge quadrupole and magnetic dipole form factors 
$F_{ch}({\bf Q}^2)$, $F_{q}({\bf Q}^2)$ and $F_{m}({\bf Q}^2)$ according to 
$A({\bf Q}^2) = F_{ch}^2({\bf Q}^2) + {\bf Q}^4/18 \ F_{q}^2({\bf Q}^2) + 
{\bf Q}^2/6M_p^2 \ F_{m}^2({\bf Q}^2)$ \cite{Arn80};
the form factors are considered in the Breit frame in which ${\bf Q}^2=-q^2$, 
$q$ being the four-momentum transfer. The charge quadrupole and the magnetic 
dipole form factors are normalized to the deuteron quadrupole moment in units 
of fm$^2$ and to the deuteron magnetic moment in units of $\mu_N$, 
respectively.

Characteristic experimental data for $A({\bf Q}^2)$ at small momentum transfers
are displayed in Fig. 1(a). They are obtained in elastic electron scattering;
the bearing of the experimental charge radius, measured by the isotope shift
of the $1s - 2s$ transition of atomic hydrogen, on $A({\bf Q}^2)$ is displayed
in Fig. 1(b). In Fig. 1, we use the standard form of a high-resolution plot 
which emphasizes experimental errors and possible inconsistencies between 
experiments. We see that $A({\bf Q}^2)$ is at low momentum transfers not 
determined with the desired accuracy. Furthermore, the data of Refs. 
\cite{Sim81} and \cite{Pla90} show opposing trends as already emphasized in 
Ref. \cite{Pla90}; 
however, the low-momentum transfer data of Ref. \cite{Sim81} have always been
considered reliable beyond any doubt, whereas those of Ref. \cite{Pla90} have
set a new standard of experimental accuracy at intermediate momentum transfers.
Thus, the discrepancies between both data sets create general worries about
the low-momentum transfer data, most important for extracting the deuteron 
charge radius $r_{ch}$.

\par
Elastic electron scattering is often simultaneously performed on deuteron 
and proton targets. The advantage of such a simultaneous measurement is 
that the corresponding ratio of deuteron and proton charge form factors, 
$F_{ch}({\bf Q}^2)/G_{Ep}({\bf Q}^2)$, can be extracted with a smaller 
systematic error. Thus, the difference $r_{ch}^2-r_p^2$, $r_p^2$ being the
ms proton charge radius, is experimentally determined with higher accuracy
than the deuteron charge radius itself. This is the reason why experimentalists
prefer to analyze their data for the rms deuteron {\it structure} radius $r_d$,
defined by
\begin{equation}
\label{rd}
r_{d}^2=(r_{ch}^2-r_p^2)-r_n^2-r_{DF}^2.
\end{equation}
The ms structure radius contains as dominant part the experimentally 
determined difference $r_{ch}^2-r_p^2$. From this one removes (i) the 
contribution of the neutron finite e.m. size, $r_n^2$ being the neutron 
ms charge radius, and (ii) the relativistic Darwin-Foldy correction to
the nonrelativistic one-nucleon charge operator, 
$r_{DF}^2=G_E^S(0)3/(4M_N^2)=3/(4M_N^2)=0.0331$ fm$^2$, where 
$G_E^S({\bf Q}^2)$ is the isoscalar charge Sachs form factor of the nucleon
and $M_N$ the isospin-averaged nucleon mass. In Eq. (\ref{rd}), $r_{ch}^2$, 
$r_{p}^2$ and $r_{n}^2$ are observables and $r_{DF}^2$ is a well-defined 
number. The rms deuteron structure radius $r_d$ can therefore be considered
an honest observable quantity of the same standing as the deuteron charge 
radius $r_{ch}$ itself;
in many experiments it is determined with even higher accuracy.
%
%
%
%
%
\goodbreak
\begin{table}
\begin{center}
\begin{tabular}[t]{| r  r | l  l | l  l  l |}  \hline
Experiment &  & $r_{ch}$ [fm]  & $r_d$ [fm] & $r_p$ [fm] & $r_n^2$ [fm$^2$] & 
$r_{ch}$ [fm]   \\  \hline 
Berard   {\it et al.} &\cite{Ber73}$^{\phantom{a}}$ & 2.1256(64) & 1.9635(45) 
& 0.805(11) & -0.1134(24) & 2.0952(60) \\ 
Akimov   {\it et al.} &\cite{Aki79}$^{\phantom{a}}$ & 2.098\phantom{0}(26)  
& 1.935\phantom{0}(28) & 0.817(\phantom{0}8)  & -0.1170(18)  & 2.080\phantom{0}(27) \\ 
Simon    {\it et al.} &\cite{Sim81}$^{a}$ & 2.1159(65)   & 1.9540(47) 
& 0.862(12) & -0.1170(18) & 2.1160(60) \\ 
Simon    {\it et al.} &\cite{Sim81}$^{b}$ & 2.1193(80)   & 1.9576(68) & 0.862(12) 
& -0.1170(18) & 2.1190(80) \\ \hline
Schmidt-Kaler {\it et al.} & \cite{Sch93}$^{\phantom{a}}$ & 2.1303(66) & 1.9685(49) 
& 0.862(12) & -0.1130(30) & 2.1303(66) \\ \hline
\end{tabular}
\caption[Exp]
{ {\bf Experimental structure radius $r_{d}$ and charge radius $r_{ch}$ 
of the deuteron.} 
Results are given for three elastic electron scattering experiments 
\cite{Sim81,Ber73,Aki79}, and for an atomic physics experiment \cite{Sch93}. 
The {\it experimental} values \cite{Sim81,Ber73,Aki79} for the structure 
radius $r_d$ as defined in Eq. (\ref{rd}) are listed in column 2; the 
{\it resulting} deuteron charge radii using the proton and neutron charge
radii employed in these references are given in columns 3 to 5. 
In column 1 we derive updated values for the deuteron charge radius 
$r_{ch}$ starting from the experimental structure radius $r_d$ of column 2. 
For this purpose we use the presently accepted value for the proton charge 
radius,  i.e., $r_p=0.862(12)$ fm \cite{Sim80} but use the same neutron 
charge radius and the standard Darwin-Foldy term, $r^2_{DF}=0.0331$ fm$^2$, 
as in the respective references \cite{Sim81,Ber73,Aki79}. Two analyses of
the data of Ref. \cite{Sim81} using $^{(a)}$ 3rd and $^{(b)}$ 4th order
polynomials to fit the ratio $F_{ch}({\bf Q}^2)/G_{Ep}({\bf Q}^2)$ are
quoted. The atomic physics experiment \cite{Sch93} gives the radius
difference $(r^2_{ch}-r^2_p)_{exp}=3.795(19)$ fm$^2$; the corresponding
values for the charge and structure radii are obtained from Eq. (3) by
using $r_p=0.862(12)$ fm \cite{Sim80}, $r_n^2=-0.113(3)$ fm$^2$ 
\cite{Kop95} and $r^2_{DF}=0.0331$ fm$^2$.
}
\end{center}
\end{table}

%
%

\par
The existing experimental values for $r_d$ and $r_{ch}$ are listed in Table 1. 
The quoted values refer to the elastic electron scattering experiments of
Refs. \cite{Sim81,Ber73,Aki79}; the experiment of Ref. \cite{Pla90} whose data
are also shown in Fig. 1 was meant to provide information on $A({\bf Q}^2)$ at
intermediate momentum transfers; the authors did not extract the charge radius.
The most recent experimental result on the deuteron charge radius has been 
obtained in an atomic isotope shift measurement \cite{Sch93} and is also given
in Table 1. We base the discussion of the paper on the results for $r_{ch}$ in
column 1 of Table 1. We call these results experimental, although they cannot 
be found in the quoted references. The values quoted there are listed in 
columns 2 to 5. We consider the structure radius $r_d$ in column 2 
experimentally well-determined. However, in Refs. \cite{Ber73} and \cite{Aki79}
the corresponding deuteron charge radii $r_{ch}$, quoted in column 5, were 
obtained from Eq. (\ref{rd}) using the proton charge radius available at that 
time. We therefore update the charge radius $r_{ch}$ by using in Eq. (3) the 
currently accepted proton charge radius \cite{Sim80}, i.e., $r_p=0.862(12)$ fm,
and the values $r_n$ and $r_{DF}$ as employed in Refs. \cite{Sim81,Ber73,Aki79}.
We note an unsatisfactory spread of results in column 1 arising from the 
analyses of the considered experiments.

\par

In addition to the charge and structure radii, theoretical papers often 
discuss the deuteron {\it matter} radius $r_m$,

\begin{equation}
\label{rm}
r_m^2={1\over 4}\int_0^{\infty} \!\!\! dr\, r^2 \left [u^2(r)+w^2(r)\right ]. 
\end{equation}
\noindent
It is defined in terms of the deuteron $S$- and $D$-state wave functions,
i.e., $u(r)$ and $w(r)$ respectively. In Eq. (\ref{rm}) $r$ stands for the
relative distance between the two nucleons; a factor $r^2$ from the volume 
element is already included in the wave functions. The matter radius $r_m$ 
is theoretically important because it shows the differences between deuteron 
wave functions, obtained from different potential models for the NN 
interaction, in a direct and pure, though integral form. The matter radius 
is conceptually interesting in the following respects:

\par
\begin{itemize}
\item[$\bullet$]
Due to the factor $r^2$ in Eq. (\ref{rm}) the matter radius weighs heavily 
the long-range part of the deuteron wave function which is dominated by 
one-pion exchange. All realistic parametrizations of the NN interaction 
have one-pion exchange tails. The corresponding matter radii typically show 
a spread of their values which is smaller than $1\%$ as demonstrated later on.
\par
\item[$\bullet$]
All realistic parametrizations of the NN interaction account for the
deuteron binding energy with high precision. Their account of the 
proton-neutron spin-triplet scattering length $a_t$ and of the asymptotic 
$S$-wave normalization $A_S$ is far less precise. However, one observes 
\cite{Kla86,Bha90,Spr90} that the variations of $a_t$ and $A_S$ are 
linearily related to those of the matter radius $r_m$. At the experimental
value of the scattering length $a_t=5.419(7)$ fm according to 
Ref. \cite{Kla86,Sto88} the matter radius becomes $r_m=1.969(4)$ fm. 
In contrast, the experimental values 
for the asymptotic normalization $A_S$ \cite{Sto88} 
are not as precise as those for the triplet scattering length $a_t$ 
and are even in conflict with each other. 
Nevertheless, they also suggest a matter radius consistent 
with $r_m=1.969$ fm, though with much larger error bars \cite{Kla86}.
Conversely, assuming the empirical correlations between 
$a_t$, $A_S$, and the matter radius $r_m$ are physically reliable,
the experimental value for the scattering length $a_t$ suggests
--through the intermediary of the matter radius--  
an asymptotic $S$-wave normalization $A_S=0.885(2)$ fm$^{-1/2}$.
\end{itemize}

\noindent
Though we prefer to discuss the model-independent deuteron charge radius 
$r_{ch}$, we admit in all fairness that the deuteron matter radius $r_m$
is conceptually important and that it may, with the help of the 
well-established empirical linear relations, even serve as a measure of
consistency in experimental data and in theoretical predictions for
characteristic deuteron observables. Those observations are the reason
why others would like to extract the deuteron matter radius also from the 
experimental electron scattering data. This is clearly impossible
in a model-in\-de\-pen\-dent way, although it has been attempted often. 
The results of these attempts are given in Table 2.
%
%
\goodbreak
\bigskip
\begin{table}
\begin{center}
\begin{tabular}[t]{| r  l | l  l | l |} \hline
Analysis &  & $r_{ch}$ [fm] &  $r_m$ [fm] & $r_{[2]}^2$ [fm$^2$]  \\ \hline
Allen   et al.   & \cite{All81} & 2.1095(218) & 1.948\phantom{0}(23) 
& 0.0000  \\  
McTavish         & \cite{McT82} & 2.1204(\phantom{0}67) & 1.956\phantom{0}(\phantom{0}5) & 0.0150  \\ 
Klarsfeld et al. & \cite{Kla86} & 2.1146(\phantom{0}56) 
& 1.950\phantom{0}(\phantom{0}3) & 0.0135  \\ 
Mustafa          & \cite{Mus93} & 2.1189(\phantom{0}52) & 1.9547(19) 
& 0.0135  \\ 
Wong          & \cite{Won94} & 2.1150(\phantom{0}52) & 1.9502(20)& 0.0143  
\\ \hline
Schmidt-Kaler et al. & \cite{Sch93} & 2.1303(\phantom{0}66) & 1.9636(49)
& 0.0208  \\ 
%
%
Pachucki et al.  & \cite{Pac94} & 2.1331(\phantom{0}78) & 1.9666(66) 
& 0.0208  \\ \hline
\end{tabular}
\caption[RMS]{ 
{\bf Experimental charge radius $r_{ch}$ and matter radius $r_m$ of the 
deuteron.}
The first five entries in column 2 are the results of various analyses
\cite{Kla86,Won94,All81,McT82,Mus93} of elastic electron scattering data
\cite{Sim81,Ber73} aiming at the matter radius $r_m$. The non-nucleonic 
correction $r^2_{[2]}$ used in these analyses is based on $\pi$-meson 
exchange {\it only}; it is listed in column 3. Our reconstruction of 
the charge radius is done according to Eq. (\ref{rch}) and given in 
column 1. To calculate $r_{ch}$ from the matter radii $r_m$ 
\cite{Kla86,Won94,All81,McT82,Mus93} we use $r_p=0.862(12)$ fm, the 
neutron charge radius of the respective references, the standard 
Darwin-Foldy contribution, $r^2_{DF}=0.0331$ fm$^2$, the spin-orbit 
correction for the Paris potential $r^2_{SO}=-0.0015$ fm$^2$, and the 
non-nucleonic contribution $r^2_{[2]}$ of column 3. In the last two rows 
the deuteron charge and matter radii are calculated from the experimental
charge radius difference $(r^2_{ch}-r_p^2)_{exp}=3.795(19)$ fm$^2$ 
\cite{Sch93} and $(r^2_{ch}-r_p^2)_{exp}=3.807(26)$ fm$^2$ \cite{Pac94} 
as determined by the isotope shift of the $1s-2s$ transition in atomic 
hydrogen. The value of Ref. \cite{Pac94} includes higher order QED effects.
We calculate the corresponding deuteron matter and charge radii using 
the presently accepted values for the nucleonic radii, i.e., 
$r_p=0.862(12)$ fm \cite{Sim80} and $r_n^2=-0.113(3)$ fm$^2$ \cite{Kop95},
$r^2_{DF}=0.0331$ fm$^2$, $r^2_{SO}=-0.0015$ fm$^2$, and the two-nucleon
contribution $r^2_{[2]}=0.0208$ fm$^2$ of this paper. The latter value 
is an average of our results for $\Delta r_{MEC}$ for the Paris and the
four Bonn potentials in Table 3. }
\end{center}
\end{table}
%
%
%
%
The difference between the model-dependent ms matter radius $r_m^2$ and 
the model-independent ms charge and structure radii, $r_{ch}^2$ and $r_d^2$,
i.e.,
\begin{equation}
\label{rch}
r_{ch}^2  =  r_{m}^2+r_p^2+r_n^2+r_{DF}^2+r_{SO}^2+r_{[2]}^2 , 
\end{equation}

\begin{equation}
\label{rdz}
r_{d}^2=r_{m}^2+r_{SO}^2+r_{[2]}^2 ,
\end{equation}

\noindent
are summarized by two model-dependent corrections. We call these two 
corrections $r_{SO}^2$ and $r_{[2]}^2$. 

\par
The {\it first} one $r_{SO}^2$ is as $r_{DF}^2$ in Eq. (\ref{rd}) a 
relativistic correction which follows from the spin-orbit part of the 
one-nucleon charge operator, i.e.,

\begin{equation}
\label{rso}
r_{SO}^2=-6(2G_M^S(0)-G_E^S(0))P_D {1\over {8M_N^2}} .
\end{equation}

\noindent
It is model-dependent, since it depends on the deuteron $D$-state probability 
$P_D$; $2G_M^S(0)-G_E^S(0)=0.76$, $G_M^S({\bf Q}^2)$ being the isoscalar 
magnetic Sachs form factor of the nucleon. For example, for the $D$-state 
probability of the Paris potential, $P_D=5.77\%$, one obtains
$r^2_{SO}=-0.0015$ fm$^2$.

\par
The {\it second} correction $r_{[2]}^2$ is formally defined by Eqs. (5) and 
(6).
As difference between observable and unobservable quantities, $r_{[2]}^2$ is 
model-dependent. Conceptually, the term $r_{[2]}^2$ contains all contributions 
to the charge operator which are considered physically important and which 
are not accounted for by its single-nucleon parts. These corrections are 
usually of two-nucleon nature; this is the reason for the subscript [2]. Thus, 
$r_{[2]}^2$ has to contain two-nucleon meson-exchange corrections $r_{MEC}^2$. 
It may contain relativistic boost corrections $r_{boost}^2$ not included by 
the standard use of nonrelativistic wave functions. It may contain 
$\Delta$-isobar contributions $r_{\Delta\Delta}^2$ in interaction 
models with explicit $\Delta$-isobar degrees of freedom which yield 
$\Delta\Delta$-components in the deuteron wave function and corresponding 
charge and current operators. It may contain additional short-ranged 
quark-exchange effects $r_{QEC}^2$. However, this list is not exhaustive given
the ingenuity of theoreticians. We therefore write symbolically
\begin{equation}
\label{rsy}
r^2_{[2]}=r^2_{MEC}+r^2_{boost}+r^2_{\Delta\Delta}+r^2_{QEC}+\ldots .
\end{equation}
\par
Nevertheless, as long as the model-dependent correction $r^2_{[2]}$ is small
compared with the deuteron matter radius $r_m$, it makes sense to apply the 
corrections $r^2_{SO}$ and $r^2_{[2]}$ to data and extract that unobservable
quantity $r_m$ from them.
The theoretical analyses of Table 2 usually combine data from several electron
scattering experiments.  They make the step from the structure radius $r_d$ 
to an "experimental" matter radius $r_m$ according to Eq. (\ref{rdz}). 
However, most authors do not simply use Eq. (\ref{rdz}), but reanalyse the 
original experimental data with some theoretical bias. The dependence of these
analyses on $P_D$ is only slight. The non-nucleonic correction $r_{[2]}^2$ is 
exclusively assumed to arise from one-pion exchange effects; the value 
$r_{[2]}^2=r_{MEC}^2=0.0135$ fm$^2$ due to calculations from Ref. \cite{Koh83}
is often used. Compared with the single-experiment analyses of Table 1 we see 
that different theoretical analyses of elastic electron scattering data, 
usually based on several experiments, yield more consistent and with time even
converging values. The deuteron matter radius $r_m$, the main result of those 
analyses, is listed in column 2 of Table 2; the resulting value of about
$1.950$ fm is in conflict, by $1\%$, with the value of $1.969$ fm suggested
by the experimental scattering length $a_t$ according to the empirical linear
relation. This discrepancy is considered serious. The analysis \cite{Spr93}
of the Saclay data \cite{Pla90} is not included in Table 2. These data do not 
extend to sufficiently low ${\bf Q}^2$ to have by themselves a major impact on
the deuteron radius. Nevertheless, Refs. \cite{Won94,Spr93} conclude that the
data set \cite{Pla90} favors a larger matter radius than the low-momentum
transfer data sets.  

In contrast to the theoretical analyses, we like to reconstruct values for the
deuteron charge radius, listed in column 1. We obtain consistent values around
$2.115$ fm for it. However, in the last two rows of Table 2 we also give our 
reinterpretation of the deuteron radius resulting from the atomic physics 
experiment \cite{Sch93} and from an analysis \cite{Pac94} of higher order QED 
corrections for it. Ref. \cite{Mar95} improves some of the corrections of 
Ref. \cite{Pac94}. A clear $1\%$ discrepancy between the deuteron charge 
radius extracted from elastic electron scattering and from the atomic physics 
experiment surfaces and has been known for some time \cite{Sau94,Mar95};
this is a discrepancy between experimental data. This discrepancy remains 
unexplained and is disturbing. In addition and to the distress of 
theoreticians, no calculation based on realistic deuteron wave functions is 
able to account for the small deuteron charge radius as obtained from the 
elastic electron scattering experiments. This failure, being also of order 
$1\%$, has been demonstrated before for the deuteron matter radius $r_m$
where
the value resulting from the electron scattering data according to Table 2 
($r_m=1.950(3)$ fm) and the value based on the experimental 
scattering length ($r_m=1.969(4)$ fm) \cite{Kla86,Mar95}, 
using the empirical linear relation of theoretical nature, 
are in serious conflict. The discrepancy between experimental data and 
theoretical predictions lead to the 
speculations of Refs. \cite{Spr90,Ker90,Won92} that quark-gluon effects 
may show up in that observable.

The theoretical analyses, quoted in Table 2, require an assumption on the 
non-nucleonic correction $r^2_{[2]}$ of the deuteron radius which is 
model-dependent. 
Its meson-exchange correction has been calculated in the past in \cite{Koh83} 
for currents of pion-range only and in rather rough approximation. Though we 
expect corrections arising from heavier mesons to be rather unimportant, we 
conclude that a new determination of meson-exchange contributions to 
$r^2_{[2]}$ is necessary and timely. 
Furthermore, because of the recent speculations on exotic effects we also 
study the importance of quark-gluon corrections in $r^2_{[2]}$ using a 
nonrelativistic quark model of the deuteron. Both calculations are done in 
Sect. 3.

\section{ Theoretical Predictions for the Deuteron Charge Radius }

\noindent
The theoretical predictions are displayed in the form

\begin{equation}
r_{ch}=[r_{m}^2+r_p^2+r_n^2+r_{DF}^2+r_{SO}^2+r_{[2]}^2]^{1\over 2},
\end{equation}

\begin{equation}
\label{rcl}
r_{ch}=[r_m^2+r_p^2+r_n^2]^{1\over 2}+{{{r_{DF}^2+r_{SO}^2+r_{[2]}^2}}\over
{2[r_{m}^2+r_p^2+r_n^2]^{1\over 2}}},
\end{equation}

\begin{equation}
\label{rde}
r_{ch}=[r_m^2+r_p^2+r_n^2]^{1\over 2}+\Delta r_{DF}+\Delta r_{SO}+
\Delta r_{[2]} .
\end{equation}

\noindent
Eq. (\ref{rcl}) exploits the smallness of the corrections $r_{DF}^2$,
$r_{SO}^2$ and $r_{[2]}^2$; the equality holds better than $0.01 \%$ and
this accuracy equals the computational accuracy. Eq. (\ref{rde}) defines
the corrections $\Delta r_{DF}$, $\Delta r_{SO}$ and $\Delta r_{[2]}$ in
an obvious way; these corrections will be recorded in this paper. We prefer
the linear form of the corrections, since the effect of contributions to
the deuteron radius can be traced in a transparent way. However, the linear
form also has a clear conceptual disadvantage: E.g., $r_{DF}^2$ is a 
model-independent number whereas 
$\Delta r_{DF} = r_{DF}^2/2[r_m^2+r_p^2+r_n^2]^{1\over 2}$
becomes model-dependent due to the denominator. Anyway, in the tables below, 
$r_{DF}^2$, $r_{SO}^2$ and $r_{[2]}^2$ can always be recovered from
the knowledge of $[r_m^2+r_p^2+r_n^2]^{1\over 2}$.

\subsection{ The Deuteron as a System of Hadrons }
\nobreak

\noindent
This subsection considers the deuteron as a system of hadrons. Only nucleon
degrees of freedom are kept active, isobar and meson degrees of freedom
are frozen into instantaneous potentials and instantaneous one- and 
two-nucleon currents. Nonrelativistic quantum mechanics is used; a 
relativistic boost correction is applied. Thus, the nontrivial correction
$r_{[2]}^2$ is assumed to arise from meson exchange and relativistic boost
corrections, i.e.,

\begin{equation}
r^2_{[2]}=r^2_{MEC}+r^2_{boost} .
\end{equation}

\par
The definition of meson-exchange currents is based on the extended
S-matrix method of Ref. \cite{Ada89}, and our use of it is described in 
Ref. \cite{Hen92}. Consistency of the meson-exchange currents with the 
underlying two-nucleon potential in form, i.e., with respect to the hadron 
form factors and with respect to the off-shell extrapolation parameters 
$(\mu , \nu)$, and in meson content is required. Calculations are carried out 
for six two-nucleon potentials, i.e., for the phenomenological Reid soft-core 
(RSC) \cite{Rei68}, for the semiphenomenological Paris \cite{Lac80} and for 
four meson-exchange Bonn \cite{Mac87} potentials. The meson-exchange currents 
used correspond to all mesons which the Bonn potentials employ, i.e., the 
pseudoscalar mesons $\pi$ and $\eta$, the vector mesons $\rho$ and $\omega$ 
and the scalar mesons $\sigma$ and $\delta$. The required isoscalar charge 
operator is expanded in powers of $(p/M_N)$, $p$ being a typical nucleon 
momentum in the nucleus, and its contributions up to the relativistic order 
$(p/M_N)^2$ are retained. The physics content of the meson-diagonal 
contributions contact/pair (pair) and retardation (ret) is explained in 
Ref. \cite{Ada89}; among the meson-nondiagonal contributions only the 
$\rho\pi\gamma$ part is kept. In this paper, we have not calculated the 
effect of {\it explicit} $\Delta \Delta$ wave function components 
\cite{Ple92} on the deuteron charge radius. Taking $\Delta \Delta$ 
components explicitly into account, Ref. \cite{Bha90} sees an increase of 
the deuteron matter radius, 
Ref. \cite{Vas81}, however, a decrease due to subsequent meson-exchange 
corrections; a further study of that cancellation is indicated. 
%
%
%
%
%
\begin{table}
\begin{center}

\begin{tabular}[t]{| r | r | r | r | r| r| r|} \hline
   &  ${\bf RSC}$     & ${\bf Paris}$   & ${\bf Bonn Q}$  
   &  ${\bf Bonn A}$  & ${\bf Bonn B}$  & ${\bf Bonn C}$ \\ \hline  \hline
$\pi$-pair   & 0.0050 & 0.0036 & 0.0043 & 0.0043 & 0.0047 & 0.0050 \\ 
$\eta$-pair  &-0.0001 &-0.0001 &-0.0001 &-0.0001 &-0.0001 &-0.0000 \\ 
$\rho$-pair  &0.0001 &0.0002 &0.0005 &0.0005 &0.0003 &0.0001 \\ 
$\omega$-pair &-0.0000 &-0.0000 &-0.0001 &-0.0001 &-0.0001 &-0.0001 \\ \hline
$\rho\pi\gamma$  &0.0004 &0.0003 &0.0001 &0.0001 &0.0002 &0.0002 \\ \hline
$\pi$-ret    &0.0002 &0.0001 &0.0001 &0.0001 &0.0001 &0.0002 \\ 
$\eta$-ret   &0.0000 &0.0000 &0.0000 &0.0000 &0.0000 &0.0000 \\ 
$\rho$-ret   &0.0000 &0.0000 &0.0001 &0.0000 &0.0000 &0.0000 \\ 
$\omega$-ret &-0.0004 &-0.0004 &-0.0004 &-0.0005 &-0.0004 &-0.0004 \\ 
$\sigma$-ret &0.0005 &0.0005 &0.0005 &0.0005 &0.0005 &0.0005 \\ 
$\delta$-ret &-0.0000 &-0.0000 &-0.0000 &-0.0000 &-0.0001 &-0.0001 \\ \hline
\hline
$\Delta r_{MEC}$ &0.0057 &0.0042 &0.0050 &0.0048 &0.0051 &0.0054 \\ \hline
\end{tabular}
\caption[MEC]{
{\bf Meson-exchange contributions to the deuteron charge radius.}
The individual contributions are given in fm for six two-nucleon 
potentials in the normalization $\Delta r$ of Eq. (\ref{rde}); they are listed 
according to the exchange process (pair or ret as defined in Ref. \cite{Ada89})
and according to the exchanged meson. The row $\rho\pi\gamma$ corresponds to
the only meson-nondiagonal process taken into account. The last row 
$\Delta r_{MEC}$ contains the sum of all contributions. The last digit quoted
in the entries is numerically unstable; however, trends reflected in the last
digit appear correct.
}
\end{center}
\end{table}
\par
Our results are collected in Tables 3 and 4. The break-down of the 
meson-exchange correction $\Delta r_{MEC}$ into individual contributions is 
given in Table 3. Their potential dependence is moderate, at most $20\, \%$. 
The dominant contribution arises from the $\pi$-contact term in pseudovector 
$\pi NN$ coupling, which is equivalent to the pair term in pseudoscalar 
coupling. Previous calculations often used contributions of pion range only.
The result mostly used is that of Ref. \cite{Koh83} which is based on the 
Paris potential. We agree with that result within $10\%$, but our inclusion of 
more processes and more mesons makes $\Delta r_{MEC}$ larger by 
about $15\, \%$.
\par
The results for the deuteron charge radius are given in Table 4. We consider
the result derived from the RSC potential unreliable mainly because the RSC 
triplet scattering length is unrealistically small. In addition, the 
construction of a consistent MEC operator for a phenomenological potential 
such as the RSC potential would require some extra effort \cite{Buc85}.
Nevertheless, we list the results of the old-fashioned RSC potential as
reference for the computational comparison with previous calculations.
The results for the other potentials show a small spread only; they cluster 
around the value $2.135$ fm. They favor the value obtained in the atomic
physics experiment as Ref. \cite{Mar95} also concludes for its calculations. 
According to the results of Refs. \cite{Bha90,Vas81} this conclusion
will remain firm even if effects of $\Delta\Delta$-components in the 
deuteron wave function were taken into account.

%
%
%
%

\begin{table}

\begin{center}

\begin{tabular}[t]{| r | r | r | r | r| r| r|} \hline
  &  ${\bf RSC}$     &${\bf Paris}$   & ${\bf Bonn Q}$  
  &  ${\bf Bonn A}$  &${\bf Bonn B}$  & ${\bf Bonn C}$ \\ \hline  
$r_m$            &1.9569&1.9717&1.9684&1.9692&1.9689&1.9675 \\ \hline
$\sqrt{r_m^2+r_p^2+r_n^2}$ &2.1118&2.1255&2.1224&2.1232&2.1229&2.1216 \\ \hline
$\Delta r_{DF}$  &0.0078&0.0078&0.0078&0.0078&0.0078&0.0078 \\ \hline
$\Delta r_{SO}$  &-0.0004&-0.0003&-0.0003&-0.0003&-0.0003&-0.0003 \\ \hline
$\Delta r_{MEC}$ &0.0057&0.0042&0.0050&0.0048&0.0051&0.0054 \\ \hline \hline
$r_{ch}$         &2.1249&2.1372&2.1349&2.1355&2.1355& 2.1345 \\ \hline
\end{tabular}
\caption[MEC]{
{\bf Contributions to the charge radius in the hadronic description of the
deuteron.} 
The contributions are given in fm for six two-nucleon potentials in the
normalization $\Delta r$ of Eq. (\ref{rde}). The boost correction 
$\Delta r_{boost}$ is kinematical in the case of the deuteron; it is also 
calculated, but turns out to be zero for all retained digits. Thus, the 
non-nucleonic correction $\Delta r_{[2]}$ is identical to $\Delta r_{MEC}$ 
given in Table 3, i.e., $\Delta r_{[2]}= \Delta r_{MEC}$.
}
\end{center}
\end{table}
\goodbreak
\subsection{ The Deuteron as a Six-Quark System  }
\nobreak 

\noindent
This section considers the deuteron as a system of six nonrelativistic quarks.
The six-quark deuteron wave function is fully antisymmetrized; it is 
obtained according to the {\it Resonating Group Method} (RGM); only the 
channel with two asymptotic nucleons is retained. 
However, even a single channel calculation leads to a deuteron wave function 
which has $\Delta \Delta$ and hidden color admixtures in the short-range
region \cite{Fae83}. This is a consequence of the Pauli principle at the 
quark level.

The calculation is based on a chiral-invariant
quark hamiltonian with instantaneous two-quark potentials; the 
interaction contains a quadratic confinement potential 
and residual interactions due to gluon-, $\pi$- and $\sigma$-exchange. 

Fig. 2 displays the quark-quark interaction;
the antisymmetrizer for the six-quark wave function is considered
as part of the displayed interaction operator. 
The antisymmetrizer acts on the product wave function of two three-quark 
clusters with nucleon quantum numbers correlated by a relative 
cluster-cluster wave function 
which is not yet antisymmetrized with 
respect to quarks belonging to different clusters.

Two models are investigated for the deuteron. In the first one, called model A,
the quark hamiltonian is taken into account in full within the RGM calculation
of Ref. \cite{Val94}. 
Compared with the traditional purely nucleonic two-nucleon potentials,
model A yields an {\it attractive} nonlocal effective two-nucleon interaction
at small relative distances which arises from $\sigma$-exchange with 
simultaneous quark interchange shown in Figs. 2(b-c) 
and 2(e-g). The second one, called model B, simplifies the dynamics by  
retaining only the {\it direct} $\sigma$-exchange between three-quark nucleon 
clusters shown in Fig. 2(d). In model B the intracluster $\sigma$-exchange 
diagram of Fig. 2(a) and the $\sigma$-exchange diagrams with simultaneous 
quark interchange of Figs. 2(b-c) and 2(e-g) are omitted; the 
attraction at small relative distances is less pronounced compared with model 
A.

The comparison of the results of models A and B shows the importance of the 
attractive nonlocality due to the $\sigma$-quark exchange diagrams of 
Figs. 2(b-c) and 2(e-g) for the deuteron charge radius. Due to the
attractive nonlocality at small distances, the deuteron wave function of 
model A, i.e., the relative wave function between nucleonic three-quark 
clusters, properly renormalized to account for the norm kernel according 
to Ref. \cite{Buc89}, shows an increased probability at small relative 
distances with respect to model B. Compared with the purely nucleonic Paris 
potential wave function, both wave functions of model A and B have a
slightly increased probability at small internucleon distances but are 
otherwise similar to that of the Paris potential. The six-quark deuteron 
wave function does not exhibit any exotic structures at small relative 
distances in contrast to the speculations of Refs. \cite{Ker90,Won92}.
\par
The charge and current of the deuteron is carried by the quarks. Sample 
processes for the charge operators are displayed in Fig. 3. In addition to the 
single-quark charge operator of Figs. 3(a-c) there
are exchange corrections of two-quark nature; they arise from gluon-, $\pi$-
and $\sigma$-exchange according to Figs. 3(d-j). 
Only gluon-and $\pi$-exchange currents are considered in the calculation. 
Currents arising from $\sigma$-exchange are not taken into account; they are 
assumed to be smaller than those of pion range according to the general 
experience with $\sigma$-meson exchange currents between hadrons. 
The quark processes 
of Fig. 3 can be resummed (i) into those of single-nucleon character, 
(ii) into those of two-nucleon $\pi$-exchange character, and (iii) into novel 
ones of quark-exchange character, called quark-exchange currents (QEC). 
Especially, 
the relative weight of these three processes is important for the discussion 
of this paper. 
The quark-exchange current is defined
as the sum of all processes calculated with the proper six-quark wave
functions minus the processes (a), (d) and (g) calculated with renormalized
wave functions. In other words, the difference between processes (a), (d) 
and (g) calculated with unrenormalized and renormalized six-quark wave 
functions has to be added to the contributions of diagrams (b), (c), (f), 
(h-j) in order to obtain the proper quark exchange current \cite{Buc89}. 
A more detailed discussion of the charge and current model is 
given in Ref. \cite{Buc89}.

\par
Our results are given in Tables 5 and 6. Table 5 gives a break-down
of the quark-exchange processes into individual contributions. Table 6 sums
the results up to the charge radius. The difference in the predictions for 
models A and B shows that the short-ranged nonlocal attraction due to 
Figs. 2(b-c) and 2(e-g) indeed leads to some redistribution of matter towards
shorter distances
which affects already the deuteron matter radius and therefore the impulse
approximation to the charge radius.
We find that the novel quark-exchange current contributions
to the deuteron charge radius amount to about $20\%$ of the traditional 
$\pi$-meson exchange contribution; they are therefore of the same size as the 
non-pionic exchange current corrections considered in Subsect. 3.1. The e.m.
QEC corrections are, however, smaller than the quark exchange effects
generating the attractive nonlocality responsible for the redistribution
of matter to smaller distances. 
%
%
%
%
%
\goodbreak
\begin{table}
\begin{center}
\begin{tabular}[h]{| r | l  l |} \hline
                    &    Model   A  &  Model  B     \\ \hline    
  impulse           &    0.0001     &  0.0001       \\ \hline
  $\pi$-pair        &    0.0003     &  0.0004       \\ \hline
  gluon             &    0.0004     &  0.0004       \\ \hline \hline
  $\Delta r_{QEC}$  &    0.0008     &  0.0009       \\ \hline
\end{tabular}
\caption[QEC]{
{\bf Quark-exchange contributions to the deuteron charge radius.} 
The contributions are given in fm for the two quark models A and B in the
normalization $\Delta r$ of Eq. (\ref{rde}).
Row 1:
impulse charge operator with quark exchange according to Figs. 3(b-c). 
Row 2: pion-pair charge operator with quark exchange according to
Figs. 3(e-f), 3(h-j). 
Row 3: gluon-pair charge operator with quark exchange according to
Figs. 3(e-f), 3(h-j). 
The last row  $\Delta r_{QEC}$ contains the sum of all contributions. 
The last digit quoted in the entries is numerically unstable; however, 
trends reflected in the last digit appear correct.
}
\end{center}
\end{table}
%
%

%
%
%
\goodbreak
\begin{table}
\begin{center}
\begin{tabular}[h]{| r | r  r |} \hline
                            &   Model   A   &  Model  B    \\ \hline    
  $r_m$                     &  1.9657      &  1.9680      \\ \hline
 $\sqrt{r_m^2+r_p^2+r_n^2}$ &  2.1199      &  2.1220      \\ \hline
  $\Delta r_{DF}$           &  0.0078      &  0.0078      \\ \hline
  $\Delta r_{SO}$           & -0.0004      &- 0.0004      \\ \hline
  $\Delta r_{MEC}$          &  0.0043      &  0.0044      \\ \hline
  $\Delta r_{QEC}$          &  0.0008      &  0.0009      \\ \hline \hline
  $        r_{ch}$          &  2.1324      &  2.1348      \\ \hline
\end{tabular}
\caption[Quark]{
{\bf Contributions to the charge radius in the six-quark description of the
deuteron.}
The contributions are given in fm for the two models A and B in the
normalization $\Delta r$ of Eq. (\ref{rde}); they are graphically displayed in
Fig. 3. The processes of Figs. 3(a) and 3(d) correspond to the conventional 
ones of a single-nucleon charge operator; the results are given in the first
four rows; the experimental values $r_p=0.862(12)$ fm \cite{Sim80} and 
$r^2_n=-0.113(3)$ fm \cite{Kop95} are used. With the small quark core radius
$b=0.5184$ fm used here the experimental proton and neutron charge radii are
not exactly reproduced by the underlying quark model; see however \cite{Buc91}.
The meson-exchange correction $\Delta r_{MEC}$ accounts for the process
of Fig. 3(g); it corresponds to the canonical pion-pair term in the
hadronic description of the deuteron. The processes of Figs. 3(b-c), 3(e-f),
3(h-j) have no counterpart in the hadronic description; they yield the novel
quark exchange contributions $\Delta r_{QEC}$; their individual contributions
are listed in Table 5.
}
\end{center}
\end{table}
With respect to the speculations that quark-gluon effects may be important
for the theoretical prediction of the deuteron charge radius, we therefore
have to conclude that our calculations do not support them.
As expected, the hadronic description of the deuteron together with the 
meson-exchange corrections of Subsect. 3.1 remains valid. Furthermore, the 
discrepancy between the deuteron charge radius as extracted from elastic 
electron scattering data and its theoretical prediction remains also in a 
quark model description of the deuteron.

\section {Conclusion }
\nobreak

\noindent
We recall: For a long time the electron scattering data of Ref. \cite{Sim81},
yielding a deuteron charge radius of about $2.115$ fm, were considered valid
beyond any reasonable doubt. In fact, the value for the proton charge radius
\cite{Sim80}, still unquestioned as standard, is derived from the same
experiment. Thus, the inability of the purely nucleonic deuteron description
to account for a deuteron charge radius of $2.115$ fm -- the theoretical
predictions clustering around $2.135$ fm are off by $1 \%$ -- was taken to
be a pure {\it theory problem}. Speculations on exotic effects surfacing in
the deuteron charge radius started to flourish, though the deuteron size is
dominated by the wave function tail, the regime in which the two nucleons
are well separated. The appearance of the electron scattering data of 
Ref. \cite{Pla90} changed that situation. In principle, the experiment of
Ref. \cite{Pla90} cannot provide very detailed information on the deuteron 
charge radius; it measures the structure function $A({\bf Q}^2)$ at 
intermediate momentum transfers, but there it is in disagreement with 
Ref. \cite{Sim81}; thus, the suspicion arose that the discrepancy between 
experimental value and theoretical prediction may be rooted in an 
{\it experimental problem}
after all. This suspicion gets now further support by the result of the
atomic physics experiment \cite{Sch93} which suggests a larger experimental
deuteron charge radius of $2.130$ fm consistent with the theoretical value.
When making this conclusion we assume that e.m. corrections beyond one-photon
exchange are -- on the level of a $1 \%$ comparison -- the same for a bound
and a scattered electron; in fact, this holds true in case of the proton for
which electron scattering and an atomic physics experiment \cite{Wei92} yield
consistent data. The recalled historic evolution of the problem
provides the background for the present paper.

This paper presents a recalculation of the deuteron charge
radius within a purely nucleonic picture of the nucleus, with
a complete set of meson-exchange corrections included 
in the charge operator. A corresponding
calculation within a six-quark picture of the deuteron yields only a tiny
additional correction of the order of $0.1 \%$
arising from processes of true quark nature, which are not accounted for in 
the hadronic description. The size of this quark correction 
is consistent with a previous rough estimate \cite{Won92}.
As expected, the hadronic description
of the deuteron radius is valid. Quark-gluon effects are very small 
for the static charge properties of a nuclear system bound as lightly as 
the deuteron.
\par
Our theoretical predictions are based on six realistic two-nucleon potentials
and on a nonrelativistic constituent quark model. All yield a value of 
about $2.135$ fm for the deuteron charge radius. Only the value for RSC is, 
with $2.125$ fm, significantly smaller; we mistrust the RSC value at this level
of precision for the reasons discussed in sect. 3.1. Our theoretical predictions
of about $2.135$ fm disagree with the experimental values of around $2.115$ fm
obtained in the analyses of elastic electron scattering data and quoted in 
Table 2. We therefore also present our predictions for the deuteron longitudinal
structure function $A({\bf Q}^2)$ at finite momentum transfers ${\bf Q}$ in 
Fig. 4 which shows how the two principal data sets \cite{Sim81} and \cite{Pla90}
are in conflict. This conflict suggests that a remeasurement of $A({\bf Q}^2)$ 
in elastic electron scattering, especially at low momentum transfers, is 
necessary. According to Fig. 4, our predictions can describe the data set of 
Ref. \cite{Pla90} which, however, has only little impact on the deuteron charge
radius, and therefore was left out from Table 2. Nevertheless, the data of 
Ref. \cite{Pla90} have a trend that favors a larger charge radius consistent 
with the atomic physics experiment as was already noticed in 
Ref. \cite{Won94,Spr93}.
This conclusion is derived from the observation that all theoretical
predictions for $A({\bf Q}^2)$ in the intermediate momentum transfer range
of Fig. 4 are ordered according to their values for the charge radius:
Experimentally, data in the intermediate momentum transfer regime provide
no information on the charge radius; theoretically, however, the underlying
hadronic force and current model seems to provide a rather rigid relation
between the e.m. properties at low and intermediate momentum transfers.
\par
Finally, we note that the theoretical prediction of this paper for $r_{ch}$ 
which includes 
a complete set of exchange current operators and quark-gluon corrections 
is in agreement with the deuteron charge radius obtained in the recent 
atomic physics measurement \cite{Sch93,Pac94}.

\par
\vskip 0.5 true cm
\noindent
{\bf Added Note} 
\par
\noindent
After submitting this paper, a reanalysis of the relevant 
data on elastic electron-deuteron scattering including 
two-photon exchange corrections appeared \cite{Sic96}. 
The newly extracted deuteron charge radius $r_{ch}=2.128(11)$ fm 
\cite{Sic96} is consistent with the atomic physics experiment and 
theoretical predictions. 
\par
\vskip 0.5 true cm
\noindent
{\bf Acknowledgement} 
\par
\noindent
The authors gratefully acknowledge helpful discussions
with I. Sick and D. W. L. Sprung. The calculations were done at Regionales
Rechenzentrum f\"ur Niedersachsen (RRZN), Hannover.
%
%

%

\vfill
\eject

\begin{figure}[htb]
\label{Fig.1a}
$$\hspace{0.2cm} \mbox{
\epsfxsize 12.0 true cm
\epsfysize 16.0 true cm
\setbox0= \vbox{
\hbox {
\epsfbox{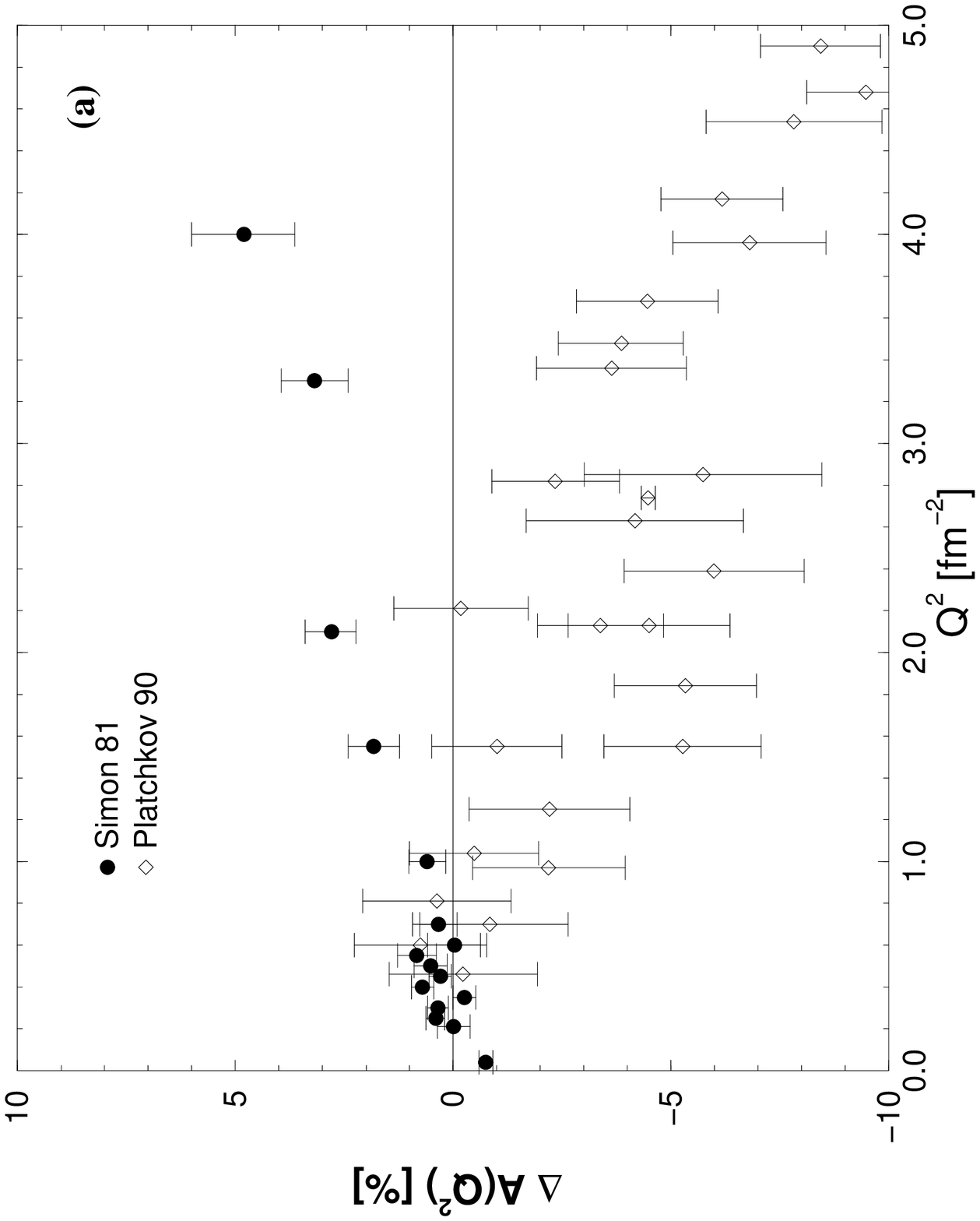}
} 
} 
\rotr0
} $$
\vspace{1.0cm}
\end{figure}

\begin{figure}[htb]
\label{Fig.1b}
$$\hspace{0.2cm} \mbox{
\epsfxsize 12.0 true cm
\epsfysize 16.0 true cm
\setbox0= \vbox{
\hbox {
\epsfbox{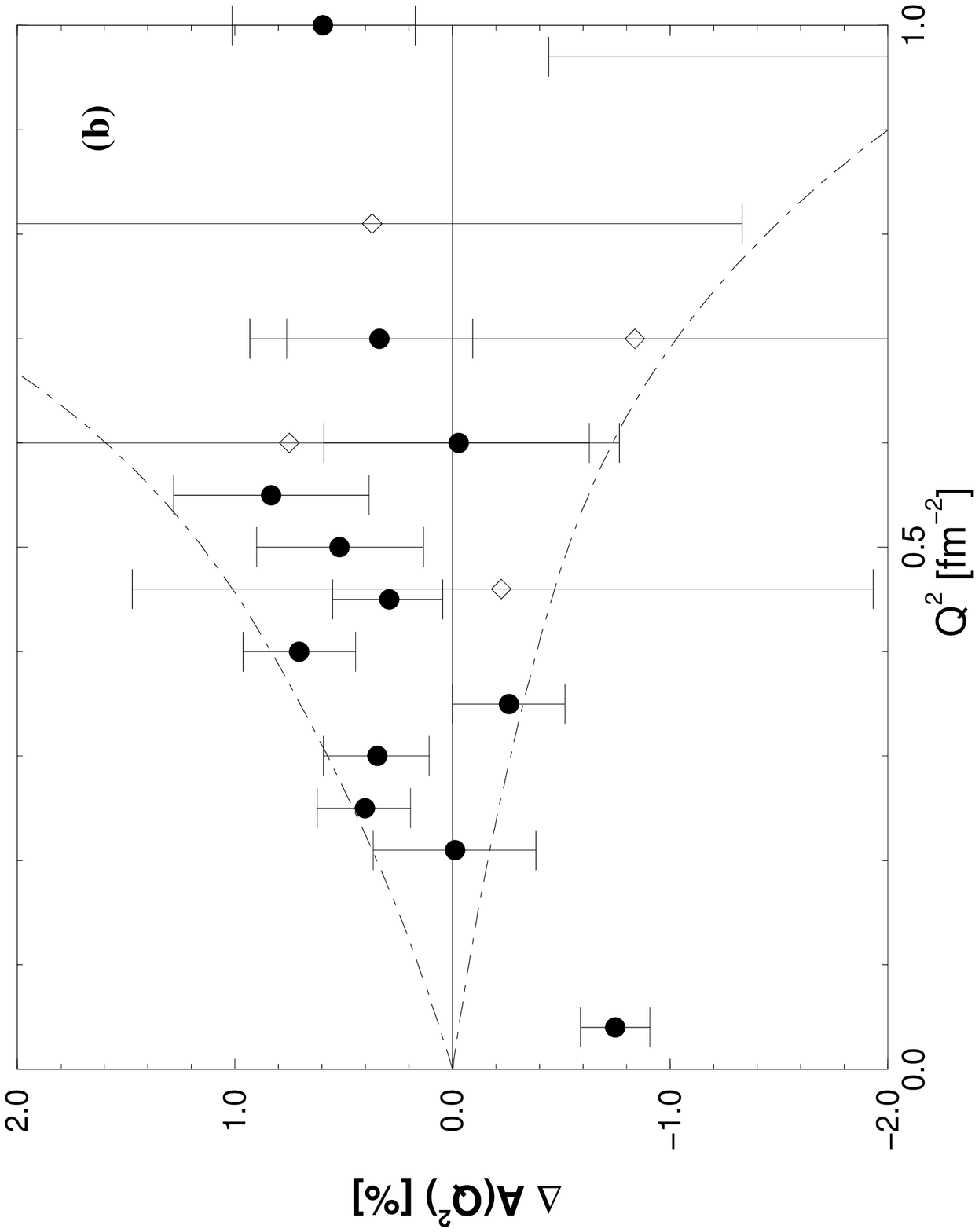}
} 
} 
\rotr0
} $$
\vspace{-1.0cm}
\caption[Fig1b]{
{\bf (a)} Experimental data for the longitudinal deuteron structure function 
$A({\bf Q}^2)$. A high-resolution representation $\Delta A({\bf Q}^2) = 
(A({\bf Q}^2) - A_{IA,Paris}({\bf Q}^2))/A_{IA,Paris}({\bf Q}^2)*100$ 
is chosen. $A_{IA,Paris}({\bf Q}^2)$ is the theoretical result derived from 
the Paris potential in nonrelativistic impulse approximation using the 
nucleonic Dirac form factor $F_1^{S}({\bf Q}^2)$ instead of the Sachs form 
factor $G_E^{S}({\bf Q}^2)$ in the nonrelativistic charge operator. The data 
are plotted in the Breit frame as a function of ${\bf Q}^2=-q^2$, $q$ being
the four-momentum transfer. The two elastic electron scattering data sets of
Refs. \cite{Sim81} (bullets) and \cite{Pla90} (diamonds) appear to be 
inconsistent; for this conclusion data up to 
${\bf Q}^2 = 5$ fm$^{-2}$ are shown.  \\
{\bf (b)} An attempt is made to indicate the importance of 
the deuteron charge
radius for the determination of the structure function $A({\bf Q}^2)$ at small
momentum transfers. The lower dashed-dotted curve refers to the linear
approximation $A({\bf Q}^2)=(1-r_{ch}^2{\bf Q}^2/6)^2$ with $r_{ch}=2.1303$ fm
from the atomic physics experiment \cite{Sch93}. The upper dashed-dotted curve
shows the corresponding linear approximation for the charge radius 
$r_{ch}=2.1146$ fm resulting from the analysis \cite{Kla86} of the elastic 
electron scattering experiments \cite{Sim81} and \cite{Ber73}. The reference 
result $A_{IA,Paris}({\bf Q}^2)$ is linearized in the same way for these 
theoretical curves. We note that the charge radius $2.1303$ fm of 
Ref. \cite{Sch93} is inconsistent with the low momentum transfer data of 
Ref. \cite{Sim81}, but seems to be consistent with the Saclay data. In the 
context of Fig. 4 we shall note that the linear approximation is poor at 
momentum transfers ${\bf Q}^2$ larger than $0.5$ fm$^{-2}$. Thus, the 
extrapolation of the elastic electron scattering data to the slope of 
$A({\bf Q}^2)$ or $F_{ch}({\bf Q}^2)$ at zero momentum transfer has to be 
done with great care. }
\end{figure}

\begin{figure}[htb]
\label{Fig.2}
$$\hspace{0.2cm} \mbox{
\epsfxsize 14.0 true cm
\epsfysize 18.0 true cm
\setbox0= \vbox{
\hbox {
\epsfbox{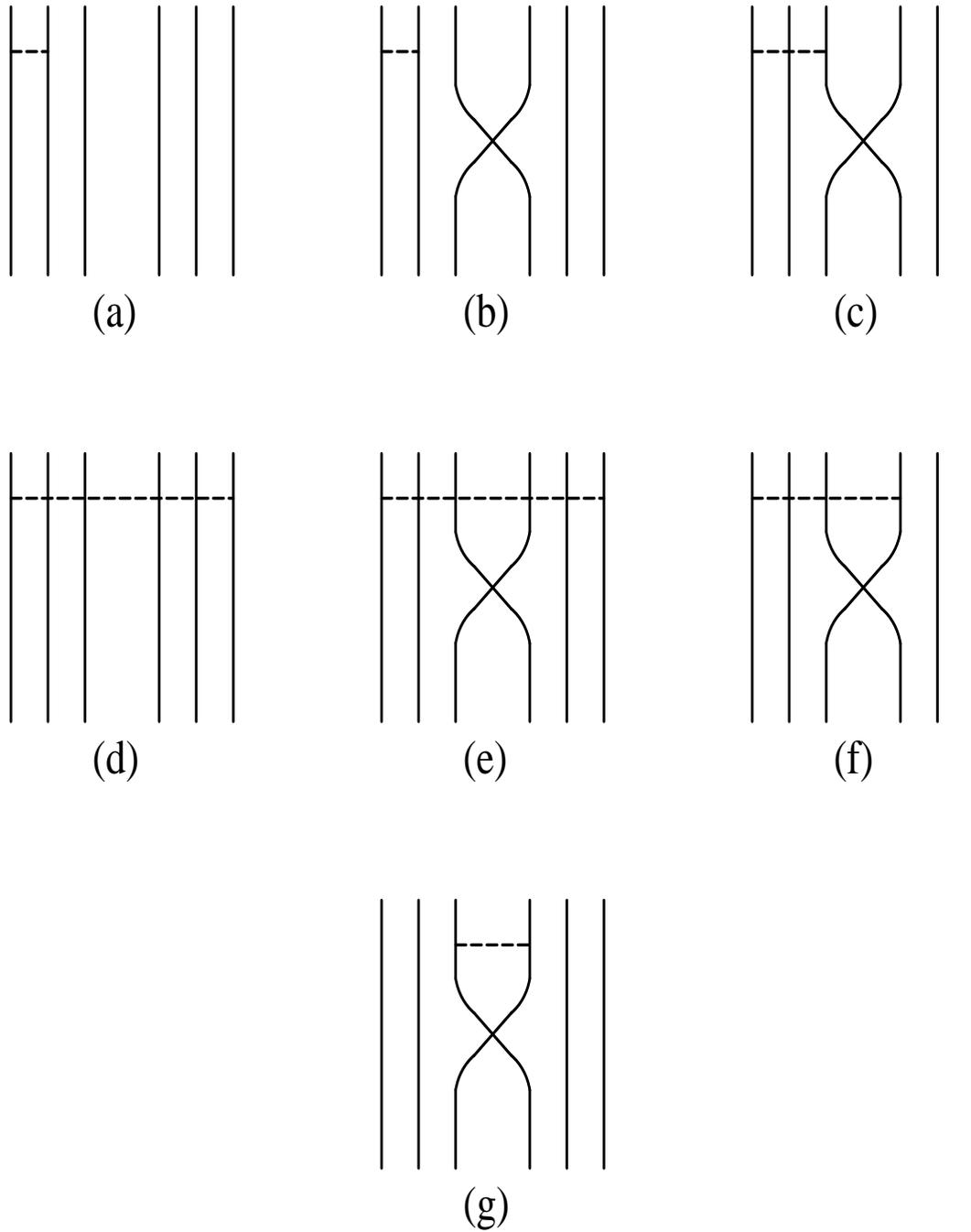}
} 
} 
\box0
} $$
\vspace{-0.75cm}
\caption[Fig.2]{
Sample contributions to the two-nucleon interaction arising from 
meson-exchange in a quark model description of the deuteron. The horizontal 
dashed line stands for the meson; $\sigma$-exchange is chosen here as an 
example. Process (a) is the intracluster term, process (d) the direct 
intercluster term and the other processes are quark-exchange terms. The 
quark model A includes all $\sigma$-exchange processes, quark model B only
the intercluster $\sigma$-exchange process (d) without quark exchange terms.
In contrast to Ref. \cite{Buc89} both models A  and B employ $\sigma$-exchange
between quarks and not between the centers of three-quark clusters.}
\end{figure}

\begin{figure}[htb]
\label{Fig.3}
$$\hspace{0.2cm} \mbox{
\epsfxsize 14.0 true cm
\epsfysize 18.0 true cm
\setbox0= \vbox{
\hbox {
\epsfbox{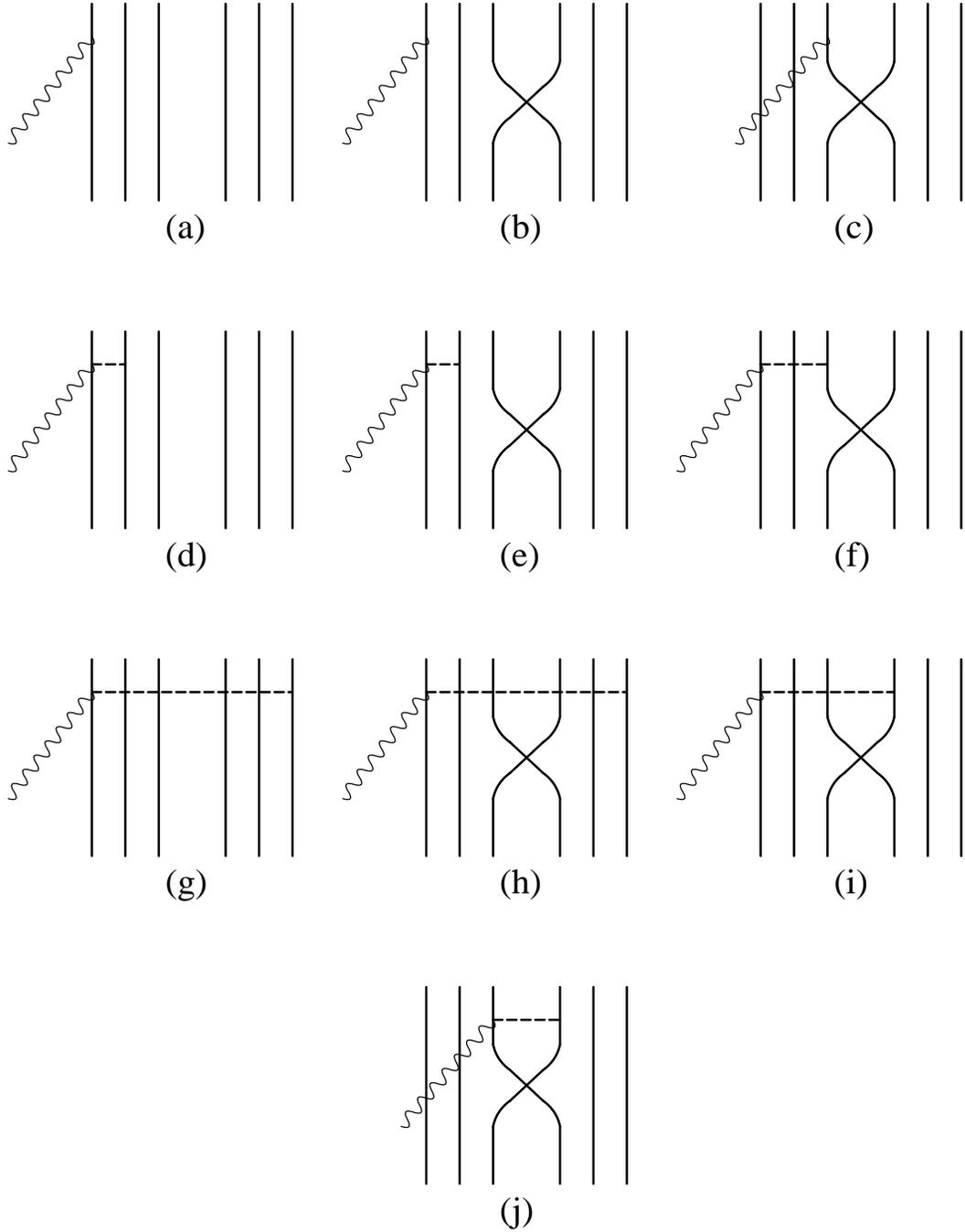}
} 
} 
\box0
} $$
\vspace{-1.1cm}
\caption[Fig.3]{ 
Sample contributions to the one-nucleon and two-nucleon currents in a quark
description of the deuteron. The current has single-quark and irreducible 
two-quark parts.
The single-quark current can be split into direct and quark-exchange pieces
according to processes (a) and (b), (c), respectively. The two-quark current
arises from $\pi$- and gluon-exchange; only the $\pi$-exchange is indicated as 
horizontal dashed line in the figure. The two-quark current can be split into  
direct intracluster (d), direct intercluster (g) and the quark-exchange 
processes (e), (f), (h)-(j). Processes (a) and (d) correspond to the 
single-nucleon current and process (g) to the two-nucleon 
pion-exchange current in a hadronic description 
when calculated with renormalized six-quark wave functions. 
The break-down of the quark-exchange processes into quark-exchange impulse, 
quark-exchange pion and quark-exchange gluon given in Table 5 is now obvious. }
\end{figure}

\begin{figure}[htb]
\label{Fig.4}
$$\hspace{0.2cm} \mbox{
\epsfxsize 16.0 true cm
\epsfysize 12.0 true cm
\setbox0= \vbox{
\hbox {
\epsfbox{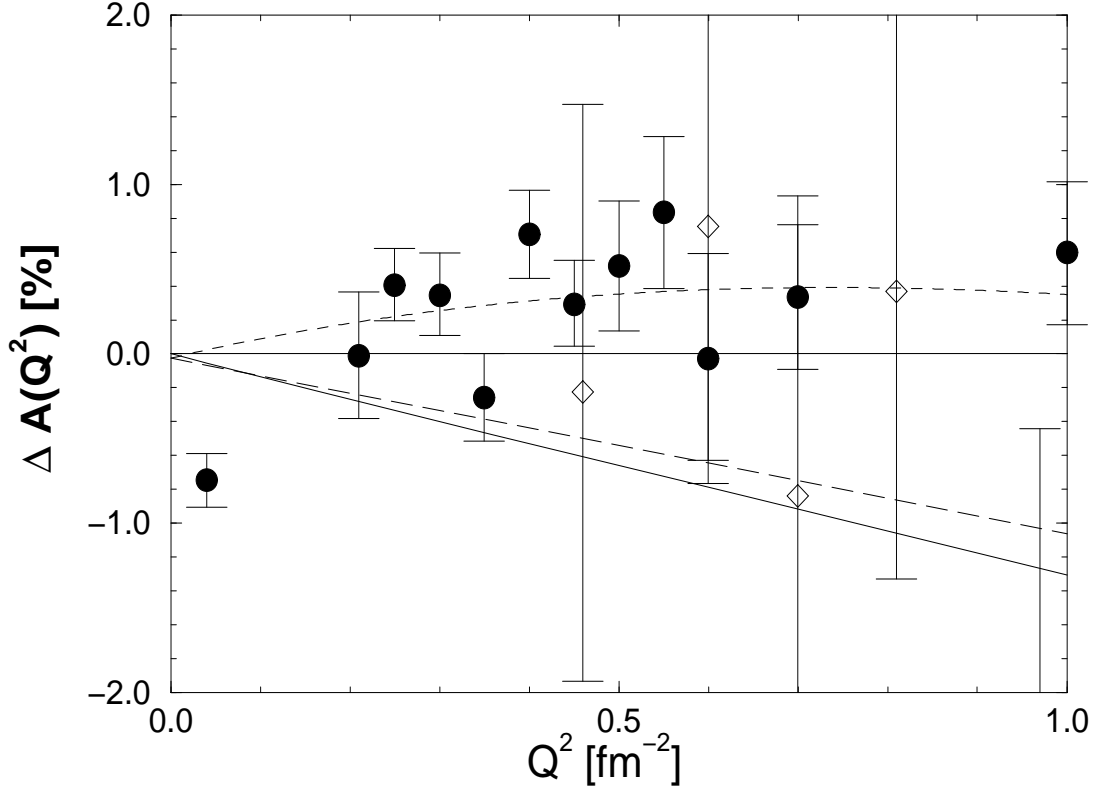}
} 
} 
\box0
} $$
\vspace{-1.0cm}
\caption[Fig.4]{
Longitudinal deuteron structure function $A({\bf Q}^2)$. A high resolution 
representation $\Delta A({\bf Q}^2) = 
(A({\bf Q}^2) - A_{IA,Paris}({\bf Q}^2))/A_{IA,Paris}({\bf Q}^2)*100$
is chosen. $A_{IA,Paris}({\bf Q}^2)$ is the theoretical result derived from the
Paris potential in nonrelativistic impulse approximation using the nucleonic 
Dirac form factor $F_1^{S}({\bf Q}^2)$ instead of the Sachs form factor 
$G_E^{S}({\bf Q}^2)$ for the nonrelativistic charge operator. Theoretical 
predictions for three realistic two-nucleon potentials RSC, Paris and Bonn C,
shown as dotted, solid, and dashed curves, respectively, are compared
with the data. The results for the Bonn A and B potentials cluster around
the Bonn C curve; our prediction for the charge radius of the Bonn C 
potential, $r_{ch}=2.1345$ fm (see table 4) is consistent with the atomic 
physics experiment (table 2). We note that the linear approximation for 
$\Delta A({\bf Q}^2)$, used in Fig. 1 to emphasize the different trend for 
$\Delta A({\bf Q}^2)$ arising from different charge radii, is not valid 
anymore for momentum transfers ${\bf Q}^2$ larger than $0.5$ fm$^{-2}$.
We observe that the theoretical predictions for the Paris and for all Bonn 
potentials cannot explain the data of Ref. \cite{Sim81}; however, they appear
to be consistent with the data set of Ref. \cite{Pla90}. }
\end{figure}

\end{document}